\documentclass[preprint]{aastex631} 

\usepackage{textcomp}  
\usepackage{comment}
\usepackage{subfigure}
\usepackage{amsmath}

\newcommand{\red}[1]{\textcolor{black}{#1}}

\newcommand{\blue}[1]{\textcolor{black}{#1}}

\received{May 6, 2021}
\revised{October 26, 2021}
\accepted{October 27, 2021}
\submitjournal{ApJ}

\shorttitle{SolarFlare2021}
\shortauthors{KamLAND collaboration}
\graphicspath{{./}{figures/}}

\begin{document}

\title{Search for Solar Flare Neutrinos with the KamLAND detector}

\correspondingauthor{N.~Kawada}
\email{kawada@awa.tohoku.ac.jp}

\newcommand{\tohoku}{\affiliation{Research Center for Neutrino Science, Tohoku University, Sendai 980-8578, Japan}}
\newcommand{\tohokuRigaku}{\affiliation{Department of Physics, Tohoku University, Sendai 980-8578, Japan}}
\newcommand{\fris}{\affiliation{Frontier Research Institute for Interdisciplinary Sciences, Tohoku University, Sendai, 980-8578, Japan}}
\newcommand{\gppu}{\affiliation{Graduate Program on Physics for the Universe, Tohoku University, Sendai 980-8578, Japan}}
\newcommand{\ipmu}{\affiliation{Institute for the Physics and Mathematics  of the Universe, The University of Tokyo, Kashiwa 277-8568, Japan}}
\newcommand{\osakarcnp}{\affiliation{Graduate School of Science, Osaka University, Toyonaka, Osaka 560-0043, Japan}}   
\newcommand{\osaka}{\affiliation{Research Center for Nuclear Physics (RCNP), Osaka University, Ibaraki, Osaka 567-0047, Japan}}
\newcommand{\tokushima}{\affiliation{Graduate School of Advanced Technology and Science, Tokushima University, Tokushima, 770-8506, Japan}}
\newcommand{\kyoto}{\affiliation{Department of Physics, Kyoto University, Kyoto 606-8502, Japan}}
\newcommand{\lbl}{\affiliation{Nuclear Science Division, Lawrence Berkeley National Laboratory, Berkeley, CA 94720, USA}}
\newcommand{\hawaii}{\affiliation{Department of Physics and Astronomy, University of Hawaii at Manoa, Honolulu, HI 96822, USA}}
\newcommand{\mituniv}{\affiliation{Massachusetts Institute of Technology, Cambridge, MA 02139, USA}}
\newcommand{\bu}{\affiliation{Boston University, Boston, MA 02215, USA}}
\newcommand{\tennessee}{\affiliation{Department of Physics and Astronomy,  University of Tennessee, Knoxville, TN 37996, USA}}
\newcommand{\tunl}{\affiliation{Triangle Universities Nuclear Laboratory, Durham, NC 27708, USA}}    
\newcommand{\chapehill}{\affiliation{The University of North Carolina at Chapel Hill, Chapel Hill, NC 27599, USA}}

\newcommand{\northcarolina}{\affiliation{North Carolina Central University, Durham, NC 27701, USA}}
\newcommand{\duke}{\affiliation{Physics Department at Duke University, Durham, NC 27705, USA}}
\newcommand{\seattle}{\affiliation{Center for Experimental Nuclear Physics and Astrophysics, University of Washington, Seattle, WA 98195, USA}}
\newcommand{\nikhef}{\affiliation{
Nikhef and the University of Amsterdam, Science Park, Amsterdam, The Netherlands}}
\newcommand{\virginia}{\affiliation{Center for Neutrino Physics, Virginia Polytechnic Institute and State University, Blacksburg, VA 24061, USA}}

\newcommand{\currentKozkov}{\affiliation{Present address: National Research Nuclear University ``MEPhI'' (Moscow Engineering Physics Institute), Moscow, 115409, Russia}}
\newcommand{\currentDima}{\affiliation{Present address: Department of Physics and Astronomy, University of Alabama, Tuscaloosa, Alabama 35487, USA}}

\newcommand{\currentHayashida}{\affiliation{Present address: Imperial College London, Department of Physics, Blackett Laboratory, London SW7 2AZ, UK}}
\newcommand{\currentUeshima}{\affiliation{Present address: National Institutes for Quantum and Radiological Science and Technology (QST), Hyogo 679-5148, Japan}}
\newcommand{\currentTakemoto}{\affiliation{Present address: Kamioka Observatory, Institute for Cosmic-Ray Research, The University of Tokyo, Hida, Gifu 506-1205, Japan}}

\author{S.~Abe}\tohoku
\author{S.~Asami}\tohoku
\author{A.~Gando}\tohoku
\author{Y.~Gando}\tohoku
\author{T.~Gima}\tohoku 
\author{A.~Goto} \tohoku
\author{T.~Hachiya}\tohoku
\author{K.~Hata} \tohoku
\author{S.~Hayashida} \altaffiliation{Present address: Imperial College London, Department of Physics, Blackett Laboratory, London SW7 2AZ, UK} \tohoku
\author{K.~Hosokawa} \tohoku
\author{K.~Ichimura} \tohoku  
\author{S.~Ieki} \tohoku
\author{H.~Ikeda}\tohoku
\author{K.~Inoue}\tohoku \ipmu 
\author{K.~Ishidoshiro}\tohoku
\author{Y.~Kamei} \tohoku
\author{N.~Kawada} \tohoku 
\author{Y.~Kishimoto} \tohoku \ipmu
\author{T.~Kinoshita} \tohoku 
\author{M.~Koga}\tohoku \ipmu 
\author{N.~Maemura}\tohoku
\author{T.~Mitsui}\tohoku
\author{H.~Miyake}\tohoku
\author{K.~Nakamura}\tohoku 
\author{K.~Nakamura}\tohoku 
\author{R.~Nakamura}\tohoku
\author{H.~Ozaki}\tohoku \gppu
\author{T.~Sakai} \tohoku 
\author{H.~Sambonsugi}\tohoku
\author{I.~Shimizu}\tohoku
\author{J.~Shirai}\tohoku
\author{K.~Shiraishi}\tohoku
\author{A.~Suzuki}\tohoku
\author{Y.~Suzuki}\tohoku 
\author{A.~Takeuchi}\tohoku
\author{K.~Tamae}\tohoku
\author{K.~Ueshima} \altaffiliation{Present address: National Institutes for Quantum and Radiological Science and Technology (QST), Hyogo 679-5148, Japan} \tohoku 
\author{Y.~Wada}\tohoku
\author{H.~Watanabe}\tohoku
\author{Y.~Yoshida} \tohoku
\author{S.~Obara}\fris
\author{A.~K.~Ichikawa} \tohokuRigaku
\author{A.~Kozlov} \altaffiliation{Present address: National Research Nuclear University ``MEPhI'' (Moscow Engineering Physics Institute), Moscow, 115409, Russia} \ipmu 
\author{D.~Chernyak} \altaffiliation{Present address: Department of Physics and Astronomy, University of Alabama, Tuscaloosa, AL 35487, USA} \ipmu 
\author{Y.~Takemoto} \altaffiliation{Present address: Kamioka Observatory, Institute for Cosmic-Ray Research, The University of Tokyo, Hida, Gifu 506-1205, Japan} \osakarcnp 
\author{S.~Yoshida}\osakarcnp
\author{S.~Umehara}\osaka
\author{K.~Fushimi}\tokushima
\author{K.~Z.~Nakamura}\kyoto
\author{M.~Yoshida}\kyoto 
\author{B.~E.~Berger}\lbl \ipmu
\author{B.~K.~Fujikawa}\lbl \ipmu
\author{J.~G.~Learned}\hawaii
\author{J.~Maricic}\hawaii
\author{S.~N.~Axani}\mituniv
\author{L.~A.~Winslow}\mituniv
\author{Z.~Fu}\mituniv
\author{J.~Ouellet}\mituniv
\author{Y.~Efremenko}\tennessee \ipmu
\author{H.~J.~Karwowski}\tunl \chapehill
\author{D.~M.~Markoff}\tunl \northcarolina
\author{W.~Tornow}\tunl \duke \ipmu
\author{A.~Li}\chapehill 
\author{J.~A.~Detwiler}\seattle \ipmu
\author{S.~Enomoto}\seattle \ipmu
\author{M.~P.~Decowski}\nikhef \ipmu
\author{C.~Grant}\bu
\author{T.~O'Donnell}\virginia
\author{S.~Dell'Oro}\virginia

\collaboration{99}{(KamLAND Collaboration)}



\begin{abstract}

We report the result of a search for neutrinos in coincidence with solar flares from the \textit{GOES} flare database.
The search was performed on a 10.8\,kton-year exposure of KamLAND collected from 2002 to 2019.
\red{This large exposure allows us to explore previously unconstrained parameter space for solar flare neutrinos.}
We found no statistical excess of neutrinos 
and established 90\% confidence level upper limits of $8.4 \times 10^7$\,cm$^{-2}$ ($3.0 \times 10^{9}$\,cm$^{-2}$) on electron anti-neutrino (electron neutrino) fluence at 20\,MeV normalized to the X12 flare, assuming that the neutrino fluence is proportional to the X-ray intensity.
\end{abstract}



\section{Introduction} \label{sec:intro}

Solar flares are the largest explosions in the solar system, releasing energy between $10^{28}$--$10^{33}$\,erg in only tens of minutes~\citep{Schrijver:2012ic}.
The mechanism of solar flares can be described as a rapid conversion of magnetic energy to thermal and kinetic energy of charged particles by reconnection of the magnetic field on the solar surface~\citep{Parker1957}. 
Observations of electromagnetic signals, ranging from radio waves to $\gamma$-rays at 100\,MeV, and neutrons emitted during solar flares contribute to the current understanding of this phenomenon~\citep{Benz2008}.

In the standard flare model, solar flares accelerate protons to more than 300\,MeV and then nuclear reactions of accelerated protons generate pions in the solar atmosphere~\citep{Hudson1995}.
Decay of these pions produces high energy (\(>\)70\,MeV) \(\gamma\)-rays and \red{MeV-scale} neutrinos.
Thus, neutrino production is expected in the standard solar flare model and the  properties of these solar flare neutrinos depend on the initial accelerated proton spectrum and flux~\citep{Kocharov1991}.

\red{
In recent decades, neutrino emission models from solar flares have been developed and such models inform the feasibility of detecting solar flare neutrinos. \citet{Fargion2004} predicted that detection of neutrinos from a large solar flare~($>10^{32}$\,erg) was feasible with Super-Kamiokande and IceCube. 
Recent updates however, predict no possibility to detect solar flare neutrinos even with Hyper-Kamiokande~\citep{Takeishi2013}. 
Another study~\citep{deWasseige:2016jvi} predicts 398--770\,$\mathrm{cm}^{-2}$ neutrino fluence at Earth in the 10--100\,MeV range, which corresponds to $\ll 1$ electron scatterings in KamLAND.
From these recent studies~\citep{Takeishi2013,deWasseige:2016jvi}, it is clear that MeV neutrino observation from a single flare is hardly feasible. 
However, by searching for a statistical excess in coincidence with a large number of solar flares, it may be possible to detect solar flare neutrinos.
Such a detection can provide an additional probe to understand the particle acceleration on the solar surface. 
}

\red{There have been several efforts to experimentally search for solar flare neutrinos. }
The Homestake experiment reported a small excess of events correlated with a large solar flare in 1991~\citep{DAVIS199413}. 
On the other hand, KAMIOKANDE II and LSD observed no excess of events associated with \red{different} solar flares~\citep{Hirata1990,Aglietta:1991fv}. 
SNO has performed a coincidence search with 842 solar flares \red{measured from 
radiation from 3\,keV to 17\,MeV with} the Reuven Ramaty High Energy Solar Spectroscopic Imager ({\it RHESSI}) and found no correlations~\citep{Aharmim2014}. 
In 2019, \red{an analysis by Borexino improved the upper limits on neutrino fluence and excluded the Homestake parameter space~\citep{AGOSTINI2021102509}.  
In this analysis, the Borexino collaboration assumed that the neutrino flux is proportional to X-ray intensity and used 472 M- and X-class solar flares selected from the {\it Geostationary Operational Environmental Satellite} ({\it GOES}) database.}
The aforementioned studies are sensitive to neutrinos in the 1--100\,MeV range. 
Recently, IceCube reported the first search for GeV-scale neutrinos related to intense $\gamma$-ray solar flares and constrained some of the parameter space associated with theoretical predictions for the neutrino flux~\citep{Abbasi:2021vxg}. 

In this paper, we present a search for solar flare neutrinos using the KamLAND data taken from 2002 March to 2019 September, which includes Solar cycle 23 and 24. 
KamLAND is a 1\,kton liquid-scintillator detector which is 
sensitive to neutrinos in the energy range between 1\,MeV and a few GeV. However, in this study we focus on 1--35\,MeV neutrinos. 
\red{For experimental studies of solar flare neutrinos, flare selection and the time window for coincidence studies are important. We discuss these in Sec.~\ref{sec:flaredata}. 
Section \ref{sec:detector} provides an overview of the KamLAND detector and the two detection channels for our solar flare neutrino search. 
The scheme of the coincidence analysis is presented in Sec.~\ref{sec:event} and \ref{sec:IBD}. 
The analysis results are converted to fluence upper limits in Sec.~\ref{sec:result}. 
}

\section{Solar flare data} \label{sec:flaredata}
\red{In a solar flare, neutrinos are generated from charged pion decay. 
Neutral pion decay emits 70--100\,MeV $\gamma$-rays. 
It is natural to identify solar flares and set the coincidence time window from the $\gamma$-ray measurements~\citep{deWasseige:2016jvi}. 
IceCube applied this strategy with the $\gamma$-ray data taken by the \textit{Fermi-LAT} satellite~\citep{Abbasi:2021vxg}. 
}

\red{
However, since the \textit{Fermi-LAT} satellite was launched in 2008, solar $\gamma$-ray burst observations are not available for the 23rd solar cycle, including the largest (class X28) flare on record that occurred on 2003 Nov 4. For this reason, we apply another strategy to identify }\red{solar flares and}
\red{the timing of particle acceleration in solar flares.
}

\blue{Hard} X-ray is an alternative channel to identify flares and set the time windows. 
\blue{Hard X-ray emission is generated from bremsstrahlung of
non-thermal electrons accelerated to relativistic velocity by a solar flare. }
\citet{2009ApJ...698L.152S} reported that there is a close proportionality between line $\gamma$-ray and hard X-ray fluence from solar flares.
The existence of line $\gamma$-rays is indirect evidence of hadronic interactions in a solar flare, 
which is one of potential sources of neutrino emission.

\red{
The light curve of solar flare X-rays differentiated by time is expected to be similar to the light curve of hard X-rays through the Neupert effect~\citep{Neupert1968,Dennis1993}.
The Neupert effect is an experimentally known effect that the derivative of soft X-ray light curves from solar flare tend to have the same timing response as microwave emission.
It can also be applied in the case of hard X-ray instead of the microwave emission.}
\red{
Thus, we can find the time window for the solar flare neutrino search using the differential soft X-ray lightcurves from \textit{GOES} satellites.
}

The advantage of \textit{GOES} soft X-ray profile compared to \textit{RHESSI} hard X-ray/$\gamma$-ray profile or \textit{Fermi-LAT} $\gamma$-ray profile is the length of the observation and the availability of stable data.
With the method mentioned above, we can use the most abundant dataset of the solar flare since 1975, which covers the 23rd and 24th solar cycles, and set the flare time window even if the hard X-ray and $\gamma$-ray observation are not available during the solar flare.
\red{This method is suggested and validated in \citet{Okamoto2020}.}

\begin{figure}[ht]
    \centering
    \includegraphics[width=0.8\linewidth]{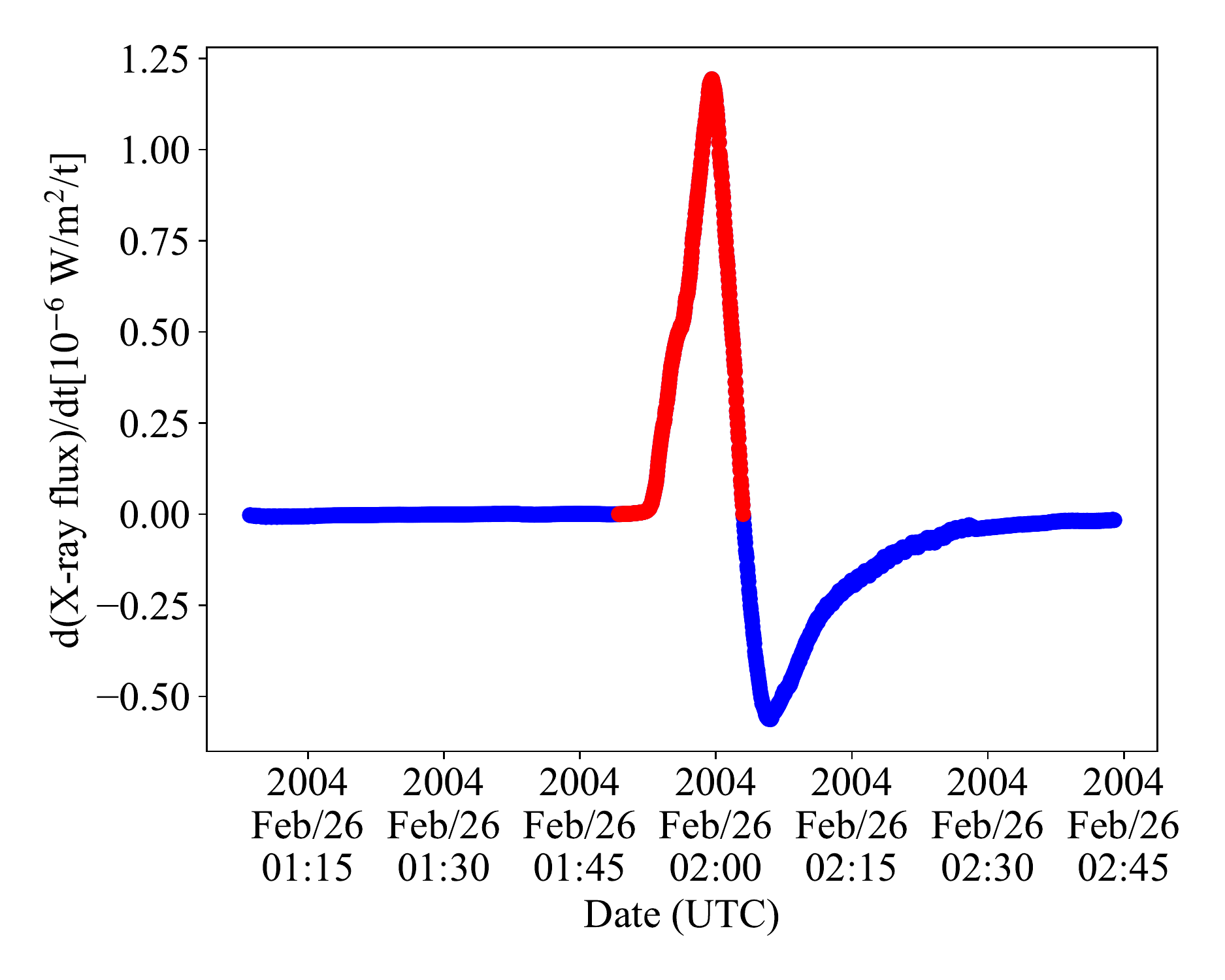}
    \caption{Derivative function of X-ray light curves in a X1.1-class flare on 2004 Feb 26. The red curve indicates the determined flare time window.}
    \label{fig:GOE040226}
\end{figure}

\red{Based on \citet{Okamoto2020}, we} determine the flare time window as follows:
(i) calculate the differential X-ray light curve, (ii) search for the peak of the differential curves,  (iii) define the time window starting from the nearest zero coefficient before the peak and ending at the nearest zero coefficient after the peak. 
\blue{The red curve in Figure~\ref{fig:GOE040226} is one of the examples for our flare time window. The duration time of this example is 1,143\,sec}.

\begin{figure}[ht]
    \centering
    \includegraphics[width=0.8\linewidth]{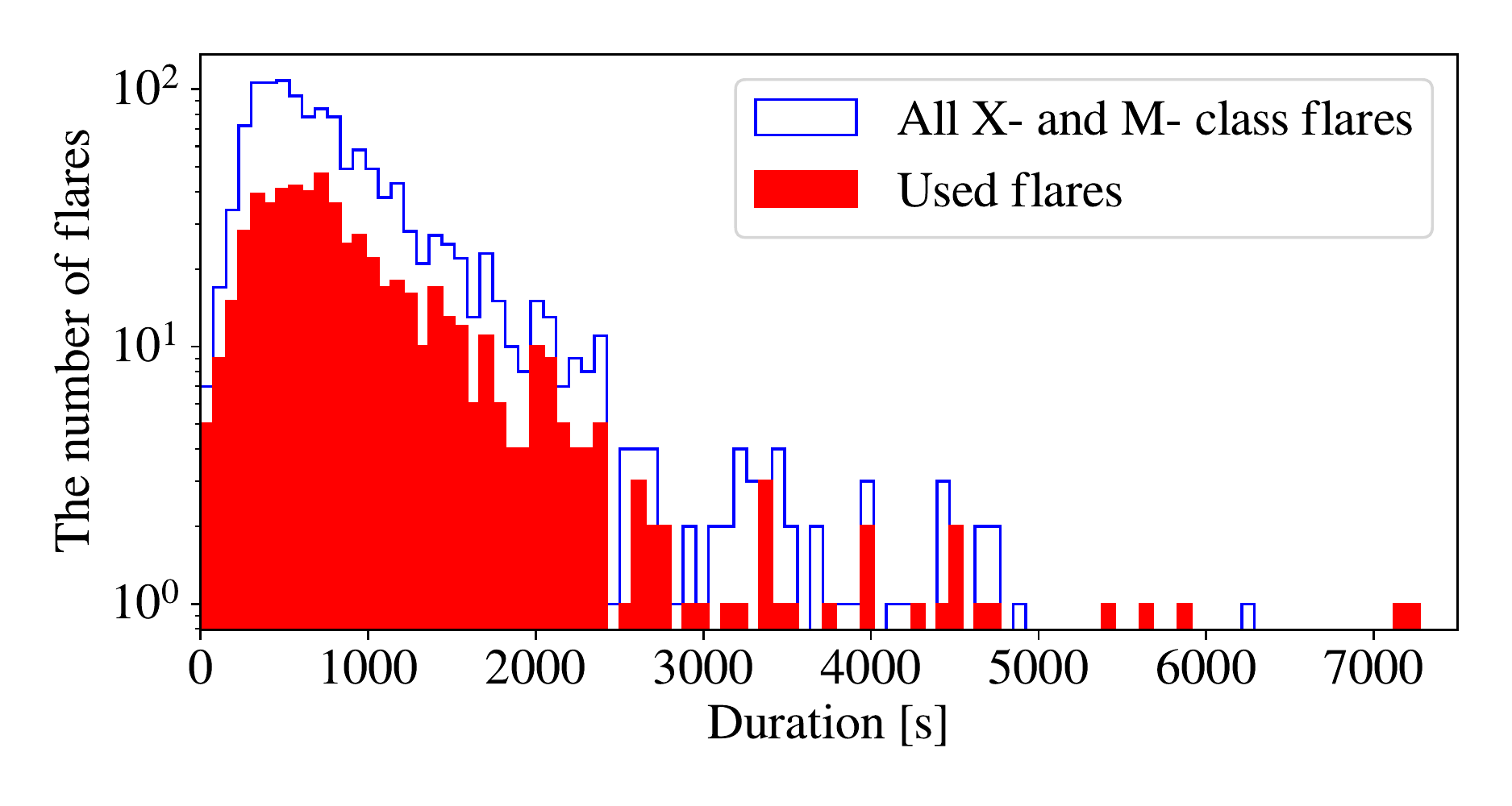}
    \caption{Distribution of the duration time of flares.}
    \label{fig:DurationDistribution}
\end{figure}

\begin{figure}[ht]
    \centering
    \includegraphics[width=0.8\linewidth]{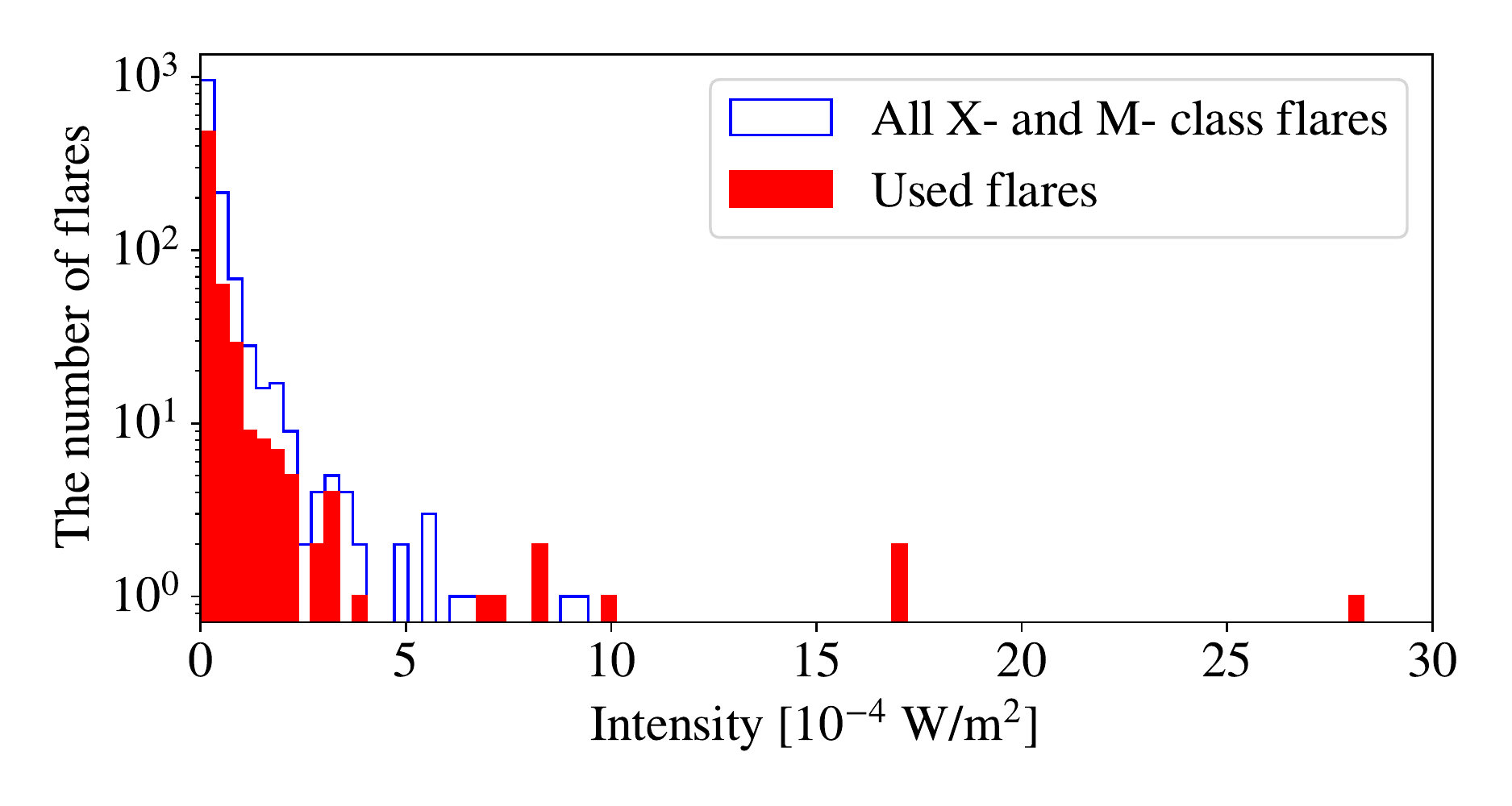}
    \caption{Distribution of X-ray intensity.}
    \label{fig:IntensityDistribution}
\end{figure}

We obtain the flare list from the {\it GOES} X-ray database \blue{at National Oceanic and Atmospheric Administration}. 
After the X- and M-class selection, which was also used in the Borexino analysis~\citep{AGOSTINI2021102509}, there were 1342 flares with a total X-ray intensity of $639.3 \times 10^{-4}$\,W/m$^2$ from 2002 March to 2019 September. 
For the coincidence analysis with the KamLAND data, all time windows were required to be in a period of operation in which the livetime to running time ratio of the detector was more than 95\%. 
The KamLAND livetime is defined as the integrated period of time that the detector was sensitive to neutrinos and includes corrections for calibration periods, detector maintenance, daily run switch, etc. 
Applying these requirements, we found 614 solar flares remained.
The distributions of the duration and intensity of these flares are shown in Figure~\ref{fig:DurationDistribution} and Figure~\ref{fig:IntensityDistribution}, respectively.
The average length of the 614 time windows is 1,028\,s. 
\blue{The duration time described in Figure~\ref{fig:GOE040226} is almost the mean value}. 
The integrated intensity is \(303.0\times10^{-4}\)\,W/m\(^2\),
which is 25 times larger than the flare coincident with the Homestake excess and 1.7 times larger than the flares used in the Borexino analysis~\citep{AGOSTINI2021102509}.


\section{KamLAND detector} \label{sec:detector} 

The KamLAND detector is a large-volume neutrino detector, which is located approximately 1\,km underground under Mt.~Ikenoyama in Kamioka, Japan.
KamLAND consists of an outer water-Cherenkov detector and an inner scintillation detector.
The water-filled outer detector (OD), housed in a 10\,m-radius \(\times\) 20\,m-high cylindrical vessel, provides shielding from external \(\gamma\)-ray backgrounds and an active muon counter. The OD was instrumented with 225 20-inch Photo Multiplier Tubes (PMTs) before a refurbishment in 2016 and 140 20-inch PMTs after the refurbishment~\citep{Ozaki:2016fmr}. 
The inner detector is a 9\,m-radius stainless steel spherical tank with 1325 17-inch PMTs and 554 20-inch PMTs mounted on the inner surface. 
The main volume of the inner detector is 1\,kton liquid scintillator supported by a 6.5\,m-radius nylon/EVOH balloon installed at the center of the stainless steel tank.
This nylon/EVOH balloon is called the outer balloon.
Outside the outer balloon is filled with non-scintillating buffer oil.
Another smaller nylon balloon for KamLAND-Zen is called the inner balloon and is described later.
The details of the KamLAND detector are described in \citet{Suzuki2014}. 


KamLAND began data taking in March 2002.
From August 2011, KamLAND started the KamLAND-Zen phase to search for the neutrinoless double-beta decay of $^{136}$Xe using a nylon balloon (inner balloon) installed at the center of the detector; this inner balloon is filled with xenon-loaded liquid scintillator~\citep{Gando2016b}. 
During the initial phase, known as KamLAND-Zen~400, which ran from August 2011 to September 2015, the inner balloon radius was 1.5\,m and the mass of xenon was about 400\,kg.  
In 2018 May, the KamLAND-Zen experiment was upgraded to the so-called KamLAND-Zen~800 phase, with an enlarged inner balloon of radius 1.9\,m and double the amount of xenon (about 800\,kg) for a higher sensitivity search~\citep{Gando2020,zencollaboration2021nylon}.
For the KamLAND-Zen periods, the regions with xenon-loaded scintillator were excluded from the effective volume for the neutrino search to suppress backgrounds from the xenon nuclei, nylon balloon, and supporting structures.

KamLAND has multiple reaction channels to detect neutrinos.
We use the following two channels, neutrino-electron elastic scattering~(ES), 
$\nu  + e^- \to \nu + e^-$,  and inverse-beta decay~(IBD), 
$\bar{\nu}_e + p \to e^+ + n$. 
ES is sensitive to all flavor of neutrinos, though 
the cross section depends on the neutrino flavor. 
This channel does not provide a measurement of the neutrino energy, though the energy of the scattered electron provides a lower bound.
IBD is sensitive only to electron anti-neutrinos above 1.8\,MeV. 
The IBD cross section is roughly ten times larger than the ES cross section. 
In addition, the IBD signal has advantages to suppress backgrounds thanks to a delayed coincidence measurement. 
The positron annihilates with an electron, emitting two 511\,keV $\gamma$-rays. The positron and two $\gamma$-rays are observed as one event called the prompt event. The incident electron anti-neutrino energy, \(E_\nu\), can be reconstructed from the prompt scintillation as \(E_\nu\simeq E_\mathrm{p}+0.8\,\mathrm{MeV}\), where \(E_\mathrm{p}\) is the energy of the prompt signal.
With \red{the mean capture time about 207\,$\mu$s}, the neutron captures on a proton~(carbon) emitting a 2.2~(4.9)\,MeV $\gamma$-ray, which is called the delayed event. 
Exploiting time-spatial correlation between the prompt and delayed events, we can observe electron type anti-neutrinos in an almost background free condition.

\section{Coincidence analysis with ES} \label{sec:event}

\subsection{Basic treatment of KamLAND data}\label{sec:basic}
Most events in KamLAND are from spallation products and decays of radioactive isotopes on the inner/outer balloons and in the liquid scintillator. 
Cosmic muons passing through the liquid scintillator generate short-lived isotopes such as \(^{8}{\rm Li}\) ($\tau=1.21$\,s) and \(^{12}{\rm B}\) ($\tau=29.1$\,ms) by spallation on carbon, which is the main component of the liquid scintillator. 
The muon events and subsequent events which occur within a veto-time window were rejected as muon-spallation related events.
The details of the spallation cuts and veto-time definitions are described in \citet{Gando_2012}.
Cosmic muon spallations also generate long-lived isotopes, \(^{10}{\rm C}\).
The beta decay of \(^{10}{\rm C}\) ($\tau=27.8$\,s) were rejected by a triple-coincidence
tag of a muon, a neutron identified by neutron-capture $\gamma$-rays and the \(^{10}{\rm C}\) decay as described in \citet{Gando2016b}.
Residual decay events from spallation products after the spallation cuts and \(^{10}{\rm C}\) veto are possible backgrounds for ES events.

To avoid backgrounds from the outer balloon and the spherical stainless-steel tank, events that were detected with $r>600$\,cm are rejected, where $r$ is the distance from the center of the detector. 
To reject background from the inner balloon and the xenon-loaded liquid scintillator, a 250\,cm radius cylinder volume in the upper hemisphere and a $r<250$\,cm volume were rejected only during the KamLAND-Zen~400/800 running periods.
One of the serious radioactive isotopes in liquid scintillator is \(^{214}\)Bi in the \(^{238}\mathrm{U}\) decay series. 
Decays of \(^{214}\)Bi to \(^{214}\)Po can contribute background events. 
Due to the short lifetime of \(^{214}\)Po, these events can be tagged by time-spatial correlation. The details of {\rm Bi}--{\rm Po} veto are described in \citet{PhysRevC.85.045504}.
Exudation decay events from the {\rm Bi}--{\rm Po} veto are another possible backgrounds for ES events.

After applying the vetoes described above, we divided the KamLAND data into 22 periods for ES studies based on the operational status of the detector and the background rate. 

\subsection{Selection criteria for ES}
\begin{figure}[ht]
    \centering
    \includegraphics[scale = 0.4]{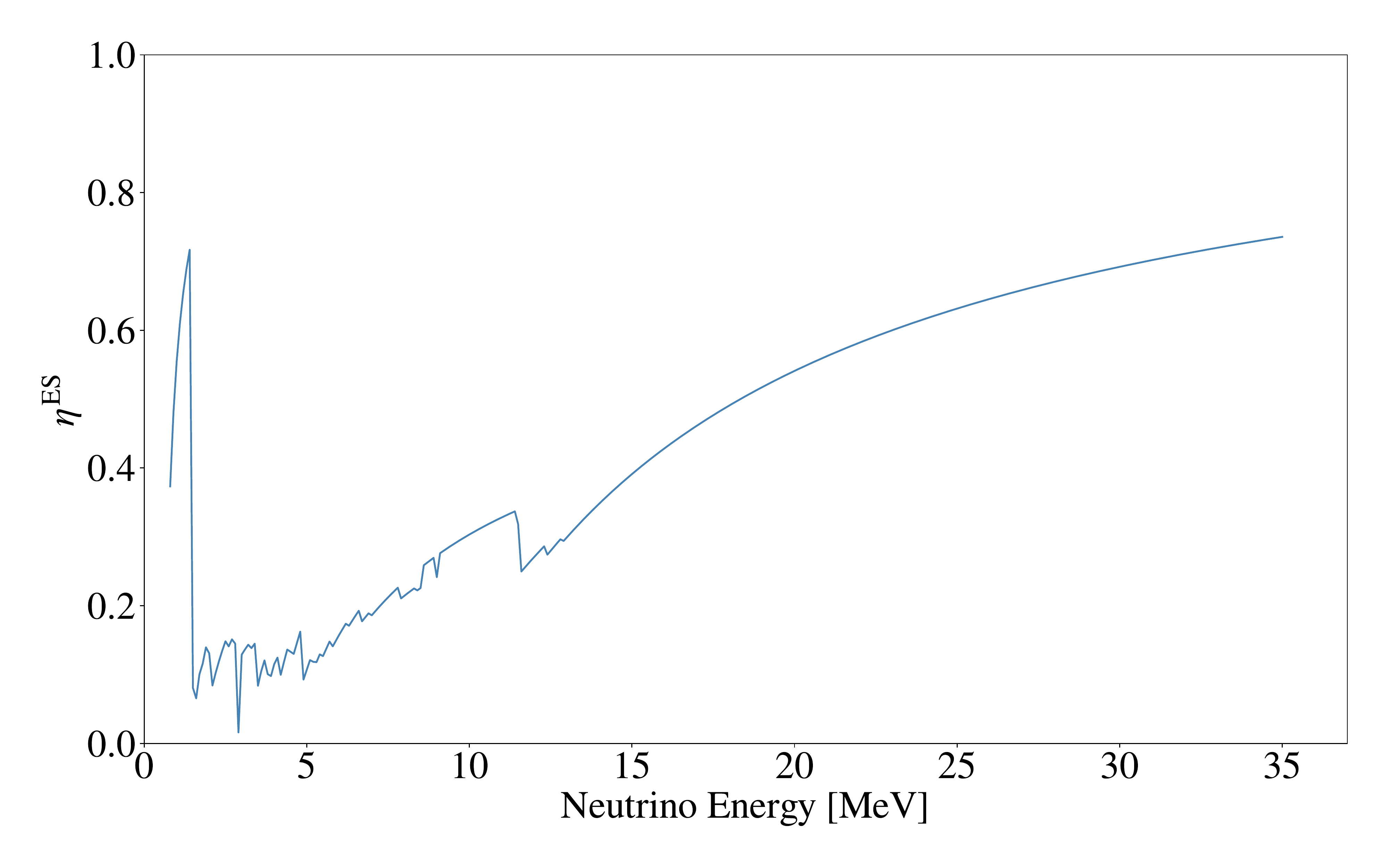}
    \caption{$\eta^\mathrm{ES}$ as a function of neutrino energy for one period. }
    \label{fig:DetectionEfficneicyForSingle}
\end{figure}

\blue{
Although there are some theoretical predictions of the spectrum of solar flare neutrinos~\citep{Kocharov1991,Fargion2004}, we} conservatively assume a monochromatic spectrum for the solar flare neutrinos, like the GRB-neutrino analysis~\citep{Super-Kamiokande:2002zkw}. 
For each assumed energy, $E_\nu$, a lower energy threshold ($E_{\textrm{th}}$) and analysis volume ($V(r_\mathrm{fid})$) were optimized to maximize the figure of merit~(FoM), defined below.
In this analysis, we used a spherical analysis volume, thus we optimized the analysis distance, $r_\mathrm{fid}$, for the volume, $V(r_\mathrm{fid})$. 
The FoM is defined as
\begin{equation}
    \textrm{FoM} = \frac{V(r_\mathrm{fid}) \times P(E_{\textrm{th}})}{\sqrt{B(r_\mathrm{fid},E_{\textrm{th}})}}, 
\end{equation}
where $P(E_{\textrm{th}}$) is the probability that the energy of the ES electron exceeds $E_{\textrm{th}}$;  $B(r_\mathrm{fid},E_{\textrm{th}})$ is the total number of background events in flare-off time of that period with $r < r_\mathrm{fid}$ and $E_{\rm th} < E_{\rm vis} < T_{\rm max}$, where $E_{\rm vis}$ is the observed energy in the KamLAND detector; $T_\mathrm{max}$ is the maximum kinetic energy of the recoil electron. 
This optimization was performed period-by-period. 
The detection efficiency, $\eta^{\rm ES} = V(r_\mathrm{fid})/V$(600\,cm)$\times P(E_{\rm th})$,
resulting from the FoM optimization considering the detector energy scale model is shown in Figure~\ref{fig:DetectionEfficneicyForSingle} as a function of the incident neutrino energy.
The shape of $\eta^{\rm ES}(E_\nu)$ depends on the vertex distribution of external \(\gamma\)-ray backgrounds penetrating the tanks from the rock surrounding the detector.

\blue{}

\subsection{Background estimation and $\chi^2$ studies}

Around $E_\nu = 3$\,MeV, the background behavior changes. 
Below 3\,MeV, there are large number of backgrounds from radioactive decays in the balloons, PMTs and the inner detector tank, such as $^{208}$Tl and $^{40}$K. 
On the other hand, contributions from radioactivity are negligible above 3\,MeV. 
Thus, we estimated the backgrounds above and below 3\,MeV separately. 

\begin{figure}[ht]
    \centering
    \includegraphics[scale = 0.6]{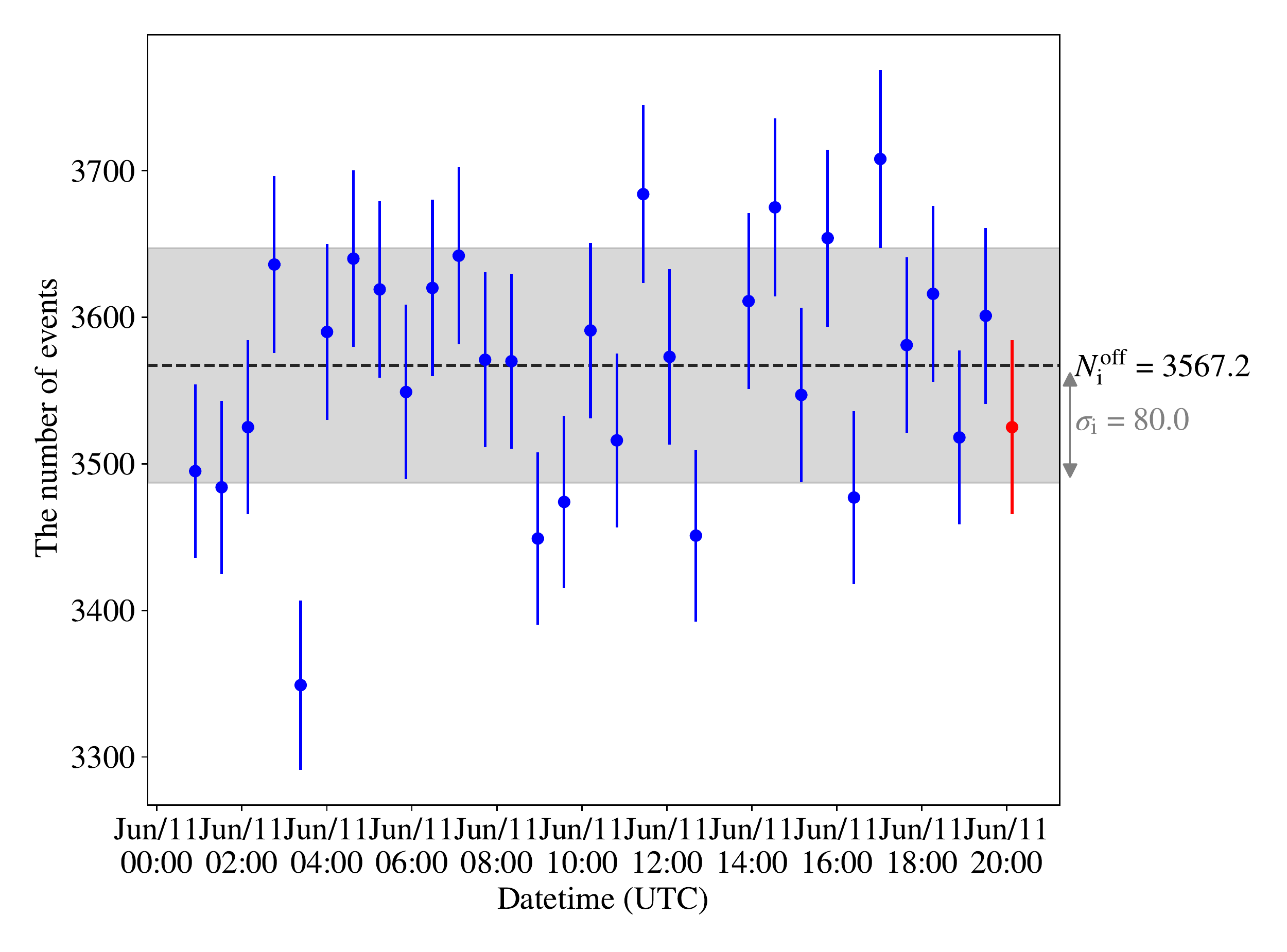}
    \caption{Example of \(N_i^{\rm on}\), \(N_i^{\rm off}\) and \(\sigma_i\) for the $i$-th flare as defined in the text. The red point is the number of observed events, \(N_i^{\rm on}\), in the flare time window. 
    The thirty blue points show the event rate in each of the off-time windows scaled by the detector livetime in that window. 
    The horizontal dashed black line shows \(N_i^{\rm off}\). The horizontal gray band shows the region [\(N_i^{\rm off}-\sigma_i\), \(N_i^{\rm off}+\sigma_i\)],  
    In this example, the flare is the M1.8 flare in 2003. Assumed neutrino energy is 1.0\,MeV. $N_i^{\rm off}$ and $\sigma_i$ are 3567.2 and 80.0, respectively.
    }
    \label{fig:NoffErrEstimation}
\end{figure}

Firstly, we describe the background determination below 3\,MeV and flare coincidence analysis. 
Since the radioactive background rate was not sufficiently stable to estimate the rate in the solar flare time (on-time) due to the liquid scintillator convection in time scale of hours, we estimated the accidental background in the following way.
For the $i$-th flare, 30 off-time windows were opened within a week before the flare.
The duration of each off-time window was the same as of the on-time window. 
In each off-time window, the number of events with $r < r_\mathrm{fid}$ and $E_{\rm th} < E_{\rm vis} < T_{\rm max}$ was counted; these are shown as blue dots in Figure~\ref{fig:NoffErrEstimation}.
We used the average ($N_i^{\rm off}$) and standard deviation ($\sigma_i$) of these off-time samples to estimate the expected number of background events with uncertainty for the associated on-time window. In Figure~\ref{fig:NoffErrEstimation}, $N_i^{\rm off}$ and $\sigma_i$ are shown as a horizontal dashed line and a gray shaded region, respectively.
Figure~\ref{fig:NoffErrEstimation} is one example from the M1.8 flare in 2003 with $E_{\rm \nu}=1.0$\,MeV, where \(E_\mathrm{th}=0.4\,\mathrm{MeV}\) and \(r_\mathrm{fid}=600\,\mathrm{cm}\). In this case, $N_i^{\rm off}$ and $\sigma_i$ are 3567.2 and 80.0, respectively. 

The expected number of events in the on-time window with the solar flare signal of the $i$-th flare is defined as $n_i \equiv N_i^{\rm BG}+w_i \eta^\mathrm{ES} \alpha^\mathrm{ES} I_i$; the first term  represents the number of background events in the on-time window and the second term corresponds to the number of signal events.
The $N_i^{\rm BG}$ are assumed to follow a Gaussian distribution with a mean of \(N_i^{\rm off}\) and a standard deviation of \(\sigma_i\). 
In the second term, $I_i$ is the flare intensity in units of [\(10^{-4}\)\,W/m\({}^2\)]; 
$\alpha^\mathrm{ES}$ is a scale factor that connects between the flare intensity and the number of ES in the 600\,cm-spherical volume, i.e., $\alpha^\mathrm{ES}$ means how many electron scatterings occur in the 600\,cm-spherical volume by a X1 flare; $\eta^\mathrm{ES}$ is the detection efficiency described above; $w_i$ is the detector livetime ratio in $i$-th on-time window.
The observed number of events with $r < r_\mathrm{fid}$ and $E_{\rm th} < E_{\rm vis} < T_{\rm max}$ in the on-time window for the \(i\)-th flare ($N_i^{\rm on}$) should follow a Poisson distribution with a mean of $n_i$. 
In the case of Figure~\ref{fig:NoffErrEstimation}, $N_i^{\rm on}$ is 3525, and is shown as a red dot.

The $\chi^2$ for all flares can be written as, 
\begin{equation}
    \chi^2 = 2\sum_{i\in {\rm flares}}[N_i^{\rm on} - n_i + n_i\ln(n_i/N_i^{\rm on})] + \sum_{i\in {\rm flares}}\left(\frac{N^{\rm BG}_{i}-N^{\rm off}_{i}}{\sigma_i}\right)^2.
    \label{eq:chi2ES}
\end{equation}
The second term in Equation~(\ref{eq:chi2ES}) is a $\chi^2$ penalty to account for the uncertainty on the background rate below 3\,MeV, using $\sigma_i$ as a conservative error.
In Equation~(\ref{eq:chi2ES}), $\alpha^\mathrm{ES}$ and $N^\mathrm{BG}_i$ are free parameters, i.e., this $\chi^2$ was minimized with respect to $\alpha^\mathrm{ES}$ and $N^\mathrm{BG}_i\,(i=1,2,\cdots,613)$.

Above 3\,MeV, the background rate is small and stable. 
The mean number of events in the on-time window is $n_i = \langle N^\mathrm{BG}_i \rangle + w_i \eta^\mathrm{ES} \alpha^\mathrm{ES} I_i$, where 
$\langle N^\mathrm{BG}_i \rangle$ is the background rate averaged over the period scaled by the coincidence window duration.
The $\chi^2$ is modified to 
\begin{equation}\label{eq:chi}
    \chi^2 = 2\sum_{i\in {\rm flares}}[N_i^{\rm on} - n_i + n_i\ln(n_i/N_i^{\rm on})] .
\end{equation}
This $\chi^2$ was minimized with respect to $\alpha$.

\begin{figure}[ht]
    \centering
    \includegraphics[scale = 0.6]{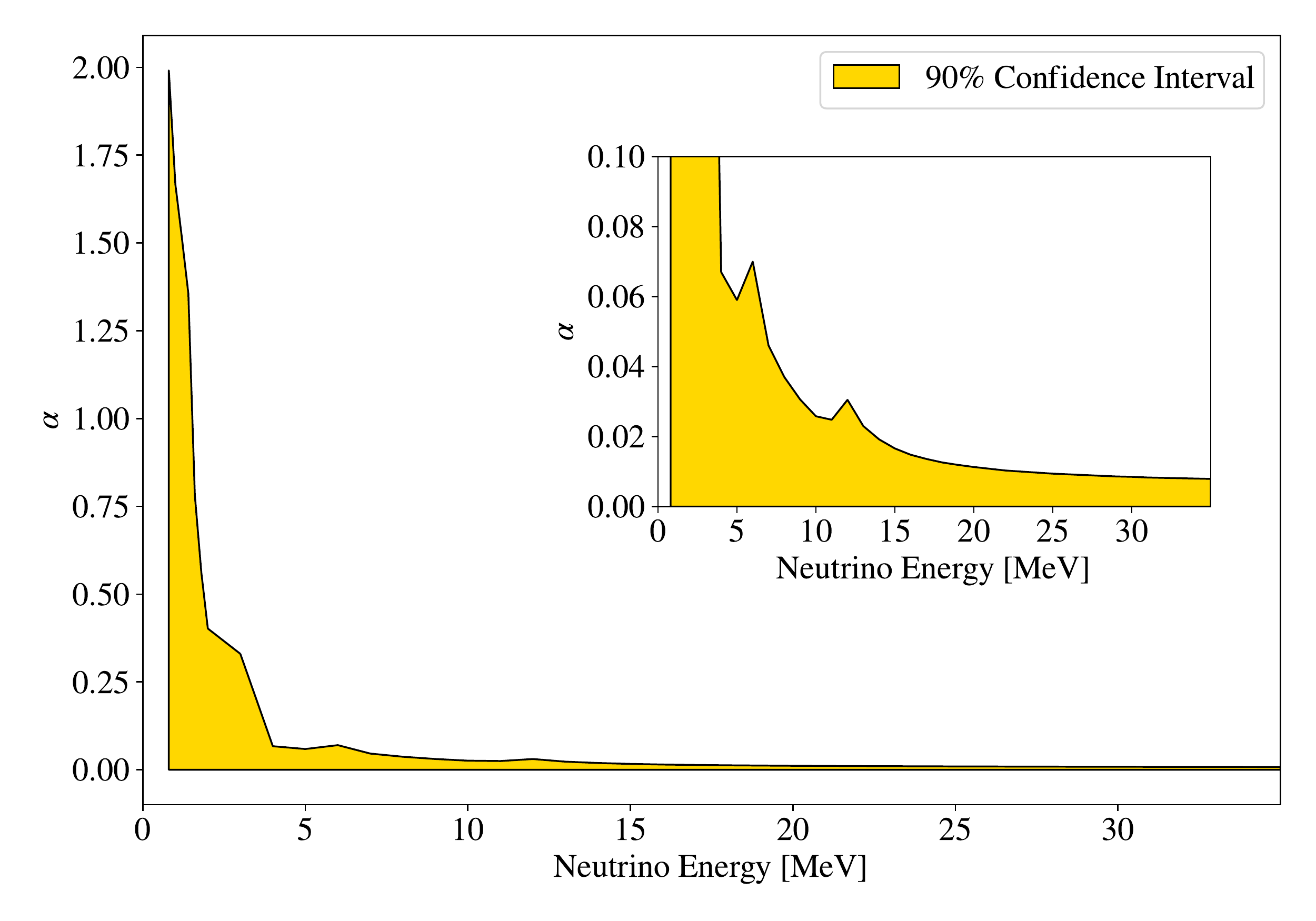}
    \caption{
    The 90\% confidence interval of \(\alpha^\mathrm{ES}\) as a function of neutrino energy in the ES analysis.
    The inset panel shows the same plot in a different vertical scale.}
    \label{PoissonChiSquareTensionForES}
\end{figure}

From the $\chi^2$ scan in our analysis range of 0.4--35\,MeV for $E_{\nu}$,
the best-fit \(\alpha^\mathrm{ES}\), $\alpha^\mathrm{ES}_\mathrm{best}$, was 0 for all assumed neutrino energies.
The 90\% confidence level~(C.L.) upper limit on $\alpha^\mathrm{ES}$, $\alpha^\mathrm{ES}_{90}$, was estimated from $\chi^2(\alpha^\mathrm{ES}_{\rm best}) + 2.7 = \chi^2(\alpha^\mathrm{ES}_{90})$.  
The 90\% confidence interval of \(\alpha^\mathrm{ES}\) is shown as a function of the assumed neutrino energy in Figure~\ref{PoissonChiSquareTensionForES}.

\blue{We have a potential problem with our time window not matching the $\gamma$-ray emission time~\citep{Okamoto2020}. 
To complement this, we also perform the coincidence analysis with a fixed time window. 
Another flare time window begins from the peak timing of the soft X-ray differential curve and runs for 1,800 seconds.
The complementary analysis shows the best fitted $\alpha^\mathrm{ES}_\mathrm{best}$ is consistent with zero within statistical errors for all assumed neutrino energies. }

\section{Coincidence analysis with IBD} \label{sec:IBD}
\subsection{Selection criteria for IBD}\label{subsec:IBDselection}
After the basic vetoes described in Section~\ref{sec:basic}, the data were divided into 12 periods. 
The definition of the periods are not the same as the ES analysis since the background conditions for IBD are quite different because of the time-spatial correlation selection. 

The prompt events were selected by requiring the reconstructed energy to be between 0.9--35\,MeV with the delayed signal on a proton (\(^{12}\mathrm{C}\)) between 1.8--2.6\,MeV (4.4--5.6\,MeV). 
The prompt-delayed pair was defined by requiring that the vertices of the two signals were less than 200\,cm apart and the time of the delayed signal must be within 0.5--1000\,\(\mu s\) of the prompt signal.
Additionally, a likelihood-based signal selection was applied to improve the purity of the IBD candidates against accidental coincidence backgrounds.

\begin{figure}[ht]
    \centering
    \includegraphics[scale = 0.4]{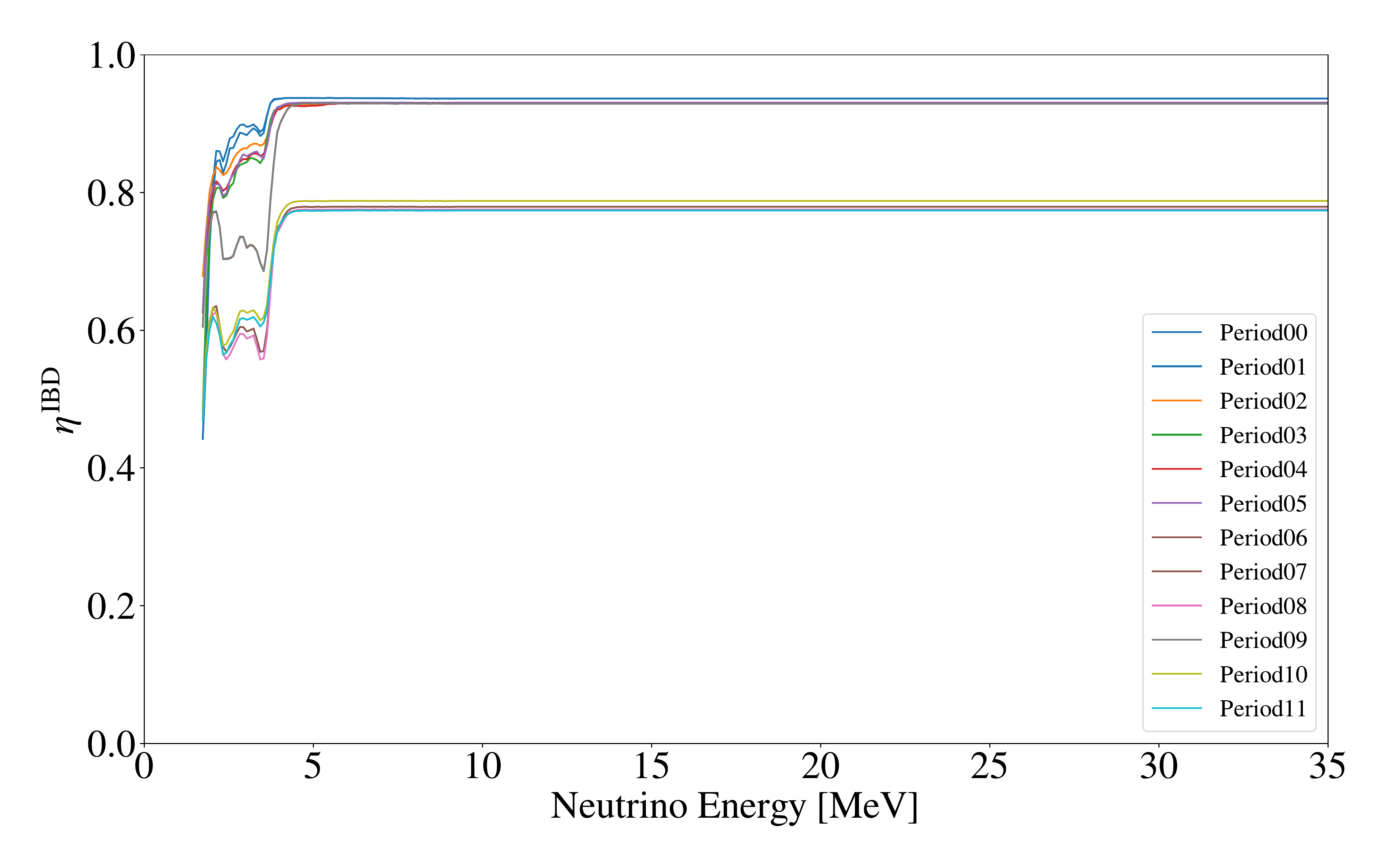}
    \caption{$\eta^\mathrm{IBD}$ as a function of neutrino energy for each periods}
    \label{fig:DetectionEfficneicyForIBD}
\end{figure}
The standard IBD candidate selection used in KamLAND, and in this analysis, is summarized in \citet{Asakura2015}.

\subsection{Background estimation and $\chi^2$ studies}
The IBD event rate is low and stable within each period because of the strong background reduction with the time-spatial correlation. 
Thus, a different $\chi^2$ is defined as 
\begin{equation}
    \chi^2 = 2\sum_{p\in {\rm period}}[N^{\rm on}_p - n_p + n_p\ln(n_p/N^{\rm on}_p)], 
\end{equation}
where $n_p = \langle N^{\rm off}_p \rangle + w_p \eta^{\rm IBD} \alpha^\mathrm{IBD} I_p$
is the expected number of events in the cumulative flare time window, i.e., summed over all coincidence windows for the flares in our sample in the $p$-th period.
The $N^{\rm off}_p$ is the expected no-flare contribution, which is estimated from the IBD event rate in the $p$-th period excluding the flare time window and scaled to the duration of the flare time window. 
$N^{\rm on}_p$ is the number of IBD events observed in the cumulative flare time window in the $p$-th period. 
$I_p$ is the cumulative X-ray intensity in the $p$-th period.
The parameter $\alpha^\mathrm{IBD}$ is a scale factor that connects between flare intensity and the number of IBD in the 600\,cm-spherical volume. 
In the IBD analysis, we used $r_{\rm fid}=600$\,cm as the analysis distance and no energy binning to count events for the $\chi^2$ study.
The $\eta^{\rm IBD}$ indicates the detection efficiency for the electron anti-neutrinos via IBD, and is computed with Monte Carlo simulation as shown in Figure~\ref{fig:DetectionEfficneicyForIBD}.
In the region below 4\,MeV, the efficiencies are reduced due to larger accidental backgrounds which affect the likelihood selection.
Because of the inner-balloon volume cuts during the KamLAND-Zen~400/800 phases as described in Section~\ref{sec:detector}, the efficiencies in some periods are lower than in other periods.
Above about 4\,MeV, the efficiencies converge to $\sim$77\% for the inner-balloon cut periods and $\sim$94\% for other periods.
$w_p$ is the detector livetime ratio.

\begin{figure}[ht]
    \centering
    \includegraphics[scale = 0.6]{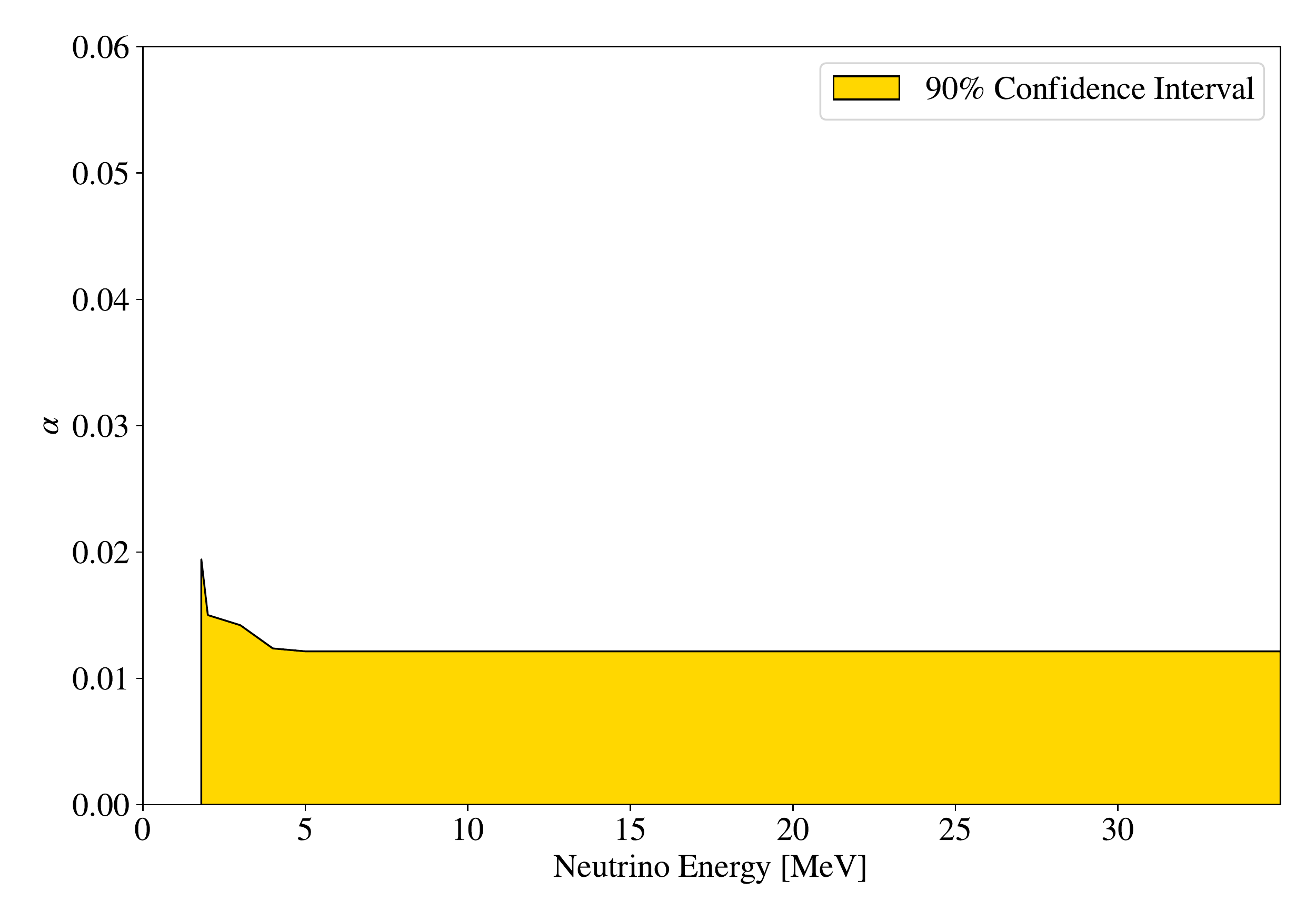}
    \caption{
    The 90\% confidence interval of \(\alpha^\mathrm{IBD}\) as a function of neutrino energy in the IBD analysis.}
    \label{PoissonChiSquareTensionForIBD}
\end{figure}

Assuming a monochromatic spectrum for the solar flare neutrinos, we varied $E_\nu$ from 1.8\,MeV to 35\,MeV, and
found $\alpha^\mathrm{IBD}$ which minimize $\chi^2$ for each assumed neutrino energies.
The best-fit values of $\alpha^\mathrm{IBD}$, $\alpha^\mathrm{IBD}_\mathrm{best}$, and the 90\% C.L. upper limits on $\alpha^\mathrm{IBD}$, $\alpha^\mathrm{IBD}_{90}$, were estimated with the same method in the ES analysis.
The $\alpha^\mathrm{IBD}_\mathrm{best}$ was 0 for all assumed neutrino energies.
The 90\% confidence interval of \(\alpha^\mathrm{IBD}\) is shown in Figure~\ref{PoissonChiSquareTensionForIBD}.   

\blue{As in the ES analysis, we also performed a fixed-time window analysis and 
obtained $\alpha^\mathrm{IBD}_\mathrm{best}=0$ for all assumed neutrino energies.}

\section{Fluence upper limit} \label{sec:result}
Although there are some theoretical predictions of the spectrum of solar flare neutrinos~\citep{Kocharov1991,Fargion2004}, it has not been experimentally measured. 
Keeping the assumption of the monochromatic signal, 
we converted \(\alpha^\mathrm{ES}_{90}\) into an upper limit on neutrino fluence, $\Phi^{\rm ES}(E_\nu)$, as
\begin{equation}
    \Phi^{\rm ES}(E_\nu) = \frac{\alpha^\mathrm{ES}_{90}(E_\nu)}{N_\mathrm{e}\int^{T_{\rm max}}_{0}\sigma(E_\nu, E_\mathrm{e})dE_\mathrm{e}}, 
\end{equation}
for the ES studies, where $N_\mathrm{e}$ is the number of electrons in the 6\,m-radius spherical volume: \(2.4\times 10^{32}\), \(E_\mathrm{e}\) is the kinetic energy of recoil electron, \(\sigma(E_\nu, E_\mathrm{e})\) is the cross section of electron scattering with the incident neutrino of energy \(E_\nu\), and \(T_{\rm max}\) is the maximum \(E_\mathrm{e}\).
For the IBD studies, the upper limit on neutrino fluence, $\Phi^{\rm IBD}(E_\nu)$, was obtained from  
\begin{equation}
    \Phi^{\rm IBD}(E_\nu) = \frac{\alpha^\mathrm{IBD}_{90}(E_\nu)}{N_p \sigma(E_\nu)}, 
\end{equation}
where \(N_p\) is the number of protons in the 6\,m-radius spherical volume: $(5.98 \pm 0.13) \times 10^{31}$, \(\sigma(E_{\nu})\) is the total cross section of IBD from \citet{Strumia_2003}. 

\begin{figure}[ht]
    \centering
    \includegraphics[scale = 0.6]{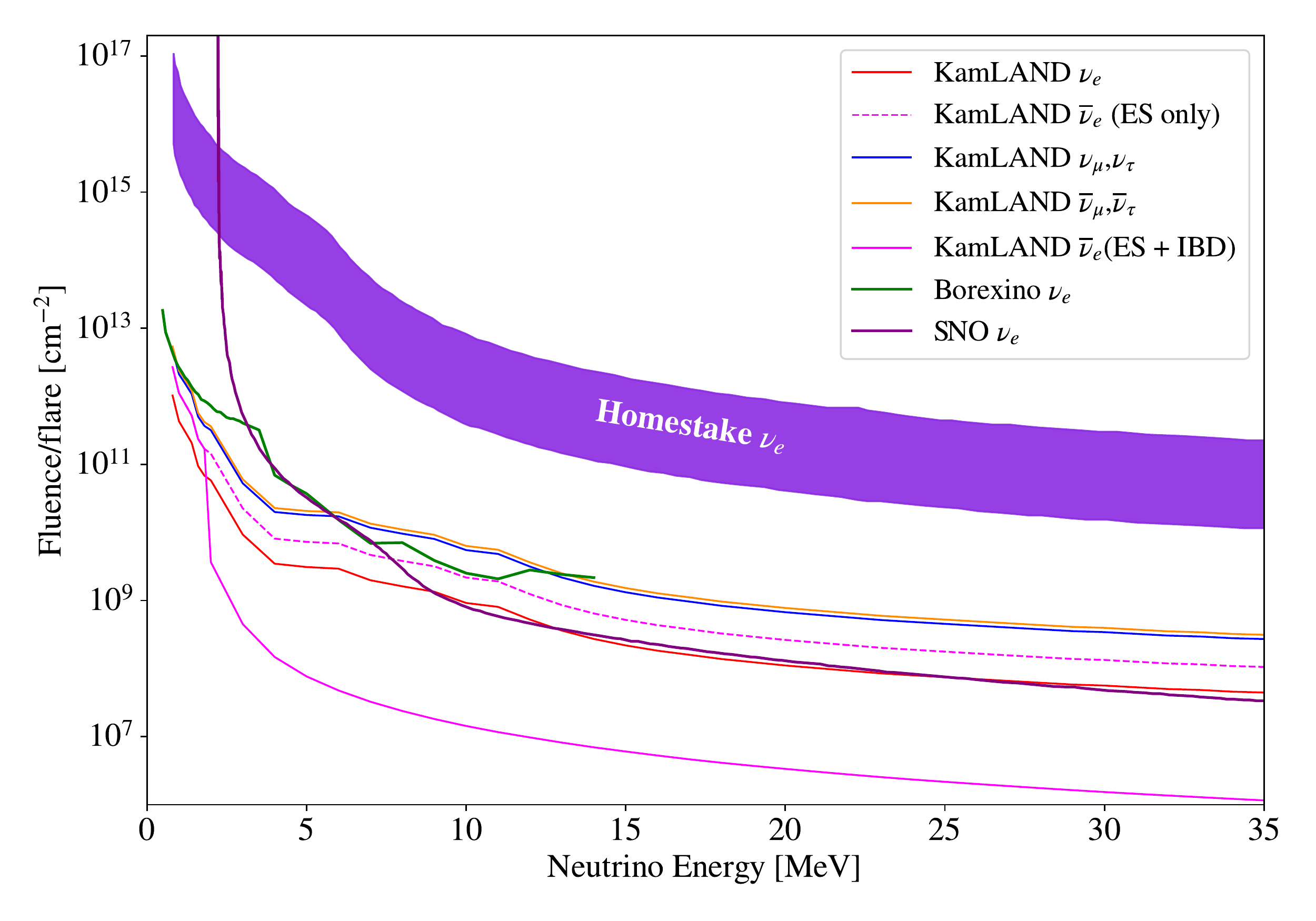}
    \caption{Fluence upper limit per flare with an assumption of equal neutrino luminosity for the flares}
    \label{fig:FluenceUpperLimitPerFlare}
\end{figure}

\red{The fluence upper limit per flare is shown in Figure~\ref{fig:FluenceUpperLimitPerFlare} with the assumption that all the flares have the same neutrino luminosity, which is discussed in SNO analysis~\citep{Aharmim2014}.}
The allowed fluence region from the Homestake excess~\citep{Aharmim2014} is shown in the purple band. 
The upper limit by SNO~\citep{Aharmim2014} and Borexino~\citep{AGOSTINI2021102509} are shown as purple and green curves, respectively.

\begin{figure}[ht]
    \centering
    \includegraphics[scale = 0.6]{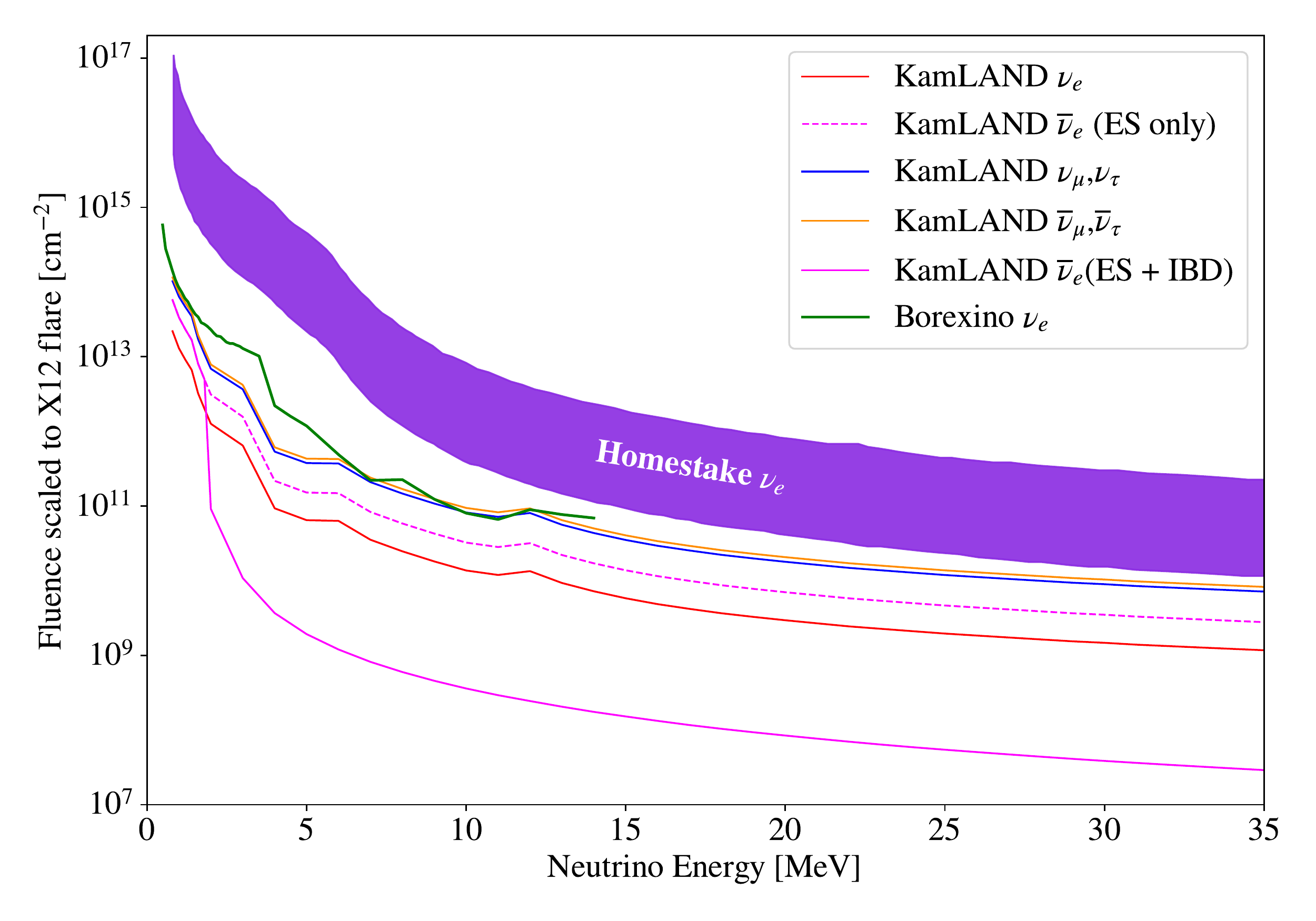}
    \caption{Fluence upper limit scaled to the Homestake flare intensity by assuming the proportionality of the neutrino luminosity to the X-ray intensity}
    \label{fig:FluenceUpperLimit}
\end{figure}

\red{Adopting another normalization that the solar flare neutrino luminosity is proportional to the X-ray intensity,}
the fluence upper limits were scaled to the Homestake flare's intensity (X12) as 
\begin{equation}
    \Phi(E_\nu)^{\rm scaled, ES/IBD} = \Phi^{\rm ES/IBD}(E_\nu) \frac{12 \times 10^{-4}\,{\rm W/m}^2}{303.3 \times 10^{-4}\,{\rm W/m}^2}, 
    \label{eq:fluencelimit}
\end{equation}
and shown in Figure~\ref{fig:FluenceUpperLimit}. 
The last term in Equation~(\ref{eq:fluencelimit}) represents the scaling factor from the flares analyzed in this work to the Homestake flare.
The upper limit by Borexino~\citep{AGOSTINI2021102509} is shown as green curve after scaling to the X12 flare.
Here, it is difficult to directly compare the results from KAMIOKANDE II~\citep{Hirata1990} and SNO~\citep{Aharmim2014} \red{in this normalization because the flare catalogue of those studies are different from this study and we do not have sufficient X-ray measurements for the corresponding flares.}

In both normalizations, the 90\% C.L. upper limits from this work exclude the entire region of the parameter space associated with the Homestake event excess for the large solar flare in 1991\red{, and are the strictest upper limits on the solar flare neutrino fluence of all flavors in the neutrino energy range of 0.4--35 MeV}.

\section{Summary and future prospect} \label{sec:conclusion}
We observe no evidence for neutrinos associated with solar flares in KamLAND.
This work places the strictest upper limits on fluence normalized to the X12 flare with the assumption that neutrino fluence is proportional to the X-ray intensity.
At 20\,MeV, the obtained 90\% C.L limits are $8.4 \times 10^7$\,cm$^{-2}$ for electron anti-neutrinos and $3.0 \times 10^{9}$\,cm$^{-2}$ for electron neutrinos. 
The Homestake region is independently rejected by this result. 
To our knowledge, this is the first time to present the upper limit normalized to the flare intensity. 
We believe that this approach is useful to compare to results from other experiments and theoretical predictions.
\red{
}
\begin{acknowledgments}
The KamLAND experiment is supported by 
JSPS KAKENHI Grants 
19H05803; 
the World Premier International Research Center Initiative (WPI Initiative), MEXT, Japan;  
Netherlands Organization for Scientific Research (NWO); 
and under the U.S. Department of Energy (DOE) Contract 
No.~DE-AC02-05CH11231,
the National Science Foundation (NSF) No.~NSF-1806440, 
NSF-2012964, 
as well as other DOE and NSF grants to individual institutions.  
The Kamioka Mining and Smelting Company has provided services for activities in the mine.  
We acknowledge the support of NII for SINET4. 
This work is partly supported by 
the Graduate Program on Physics for the Universe (GP-PU), 
and the Frontier Research Institute for Interdisciplinary Sciences, Tohoku University.
A part of this study was carried out by using the computational resources of the Center for Integrated Data Science, Institute for Space-Earth Environmental Research, Nagoya University through the joint research program.  
\end{acknowledgments}
\bibliography{main}{}
\bibliographystyle{aasjournal}



\end{document}